# Pairing-Friendly Elliptic Curves: Revisited Taxonomy, Attacks and Security Concern

Mahender Kumar, Satish Chand
School of Computer & System Sciences
Jawaharlal Nehru University, New Delhi, India
mahendjnu@gmail.com, Schand@mail.jnu.ac.in

**Abstract**: Major families of pairing-friendly elliptic curves, including BN, BLS12, BLS24, KSS16, and KSS18 have recently been vulnerable to number field sieve (NFS) attacks. Due to the recent attacks on discrete logs in $F_{q^k}$, selecting such curves became relevant again. This paper revisited the topic of selecting pairing-friendly curves at different security levels. First, we expanded the classification given by Freeman et al. [1] by identifying new families that were not previously mentioned, such as a complete family with variable differentiation and new sparse families of curves. We discussed individual curves and a comprehensive framework for constructing parametric families. We estimated the security and assessed families of the pairing-friendly curve to discover families of curves better than BN, KSS, and BLS in terms of the required key size. We also evaluated the complexity of the optimal ate pairing that has never been discussed before, except by Barbulescu et al. [2]. We demonstrated that the recent attack (TNFS) on pairing needs to increase the key size. We compared families of curves in the context of key size and selected a suitable alternative to an elliptic curve.

## 1. Introduction

The discrete logarithm problem (DLP) is the foundation of constructing public-key cryptography [3]. However, few methods such as index calculus, number field sieve (NFS), and the function field sieve (FFS) efficiently compute DLP on the finite field of the form $F_q$, where $q$ is a prime ($p$) or q = $p^k$ with $p$ is small prime and $k$ is significant, in sub-exponential time. Koblitz [4] and Miller [5] created a new group consisting of points on the elliptic curve E(F q) on the finite field $F_q$. Operations on these groups are efficiently computable, although DLP on elliptic curves (ECDLP) is difficult. According to Menezes et al. [6], each elliptic curve may have a different level of protection. They used Weil coupling ($e_r$) to reduce the DLP in the finite filed $F_{q^k}$ from the DLP in the divisor class group $E(F_q)$ using any of the sub-exponential methods for small embedding degrees $k$ can be easily solved so that $E(e_r) \subseteq E(F_{q^k})$, where prime order subsets r = # < P >, $P \in E(F_{q^k})$ and r|$q^k - 1$ (MOV reduction). They determined that the supersingular curves for $k \leq 6$ satisfy the above two criteria. Later, Frey and Ruck [7] introduced a less restrictive Tate pairing that was easily computable.

Pairing-friendly elliptic curves are those with a large prime-order subgroup r and a small embedding degree k, the two main components for building pairing-based cryptographic systems (PF-EC). In short, $k$ transforms an instance of a DLP on an elliptic curve $E(F_q)$ into an instance of a DLP on finite field elements $F_{q^k}$. A family of pairing-friendly elliptic curves (PF-EC) must satisfy the following conditions: 1) $r$ is large enough so that it is hard to solve DLP by Pollard's rho method, 2) $k$ is big enough so that subexponential methods for computing DLP in $F_{q^k}$ is equally hard as in $E(F_q)$, and 3) $k$ is relatively small enough so that the arithmetic (pairing) in $F_{q^k}$ is easily computable. The $\rho$-value defines the relationship between the two parameters, stating that the order of the subgroup r is close to the order of $E(F_q)$.

Barreto et al. [8] proposed that at the 128-bit security level, the (BN) elliptic curve with $k = 12$ is the best choice, which was recognised as a significant breakthrough in pairing-based encryption. Menezes et al. [9] created supersingular curves with embedding degrees $k \in \{2, 3, 4, 6\}$ acceptable even for pairing-based cryptosystems, although they ran into some deployment challenges. The deployment of supersingular curves is possible on the smaller characteristic $K \neq 2$ or 3. Miyaji et al. [10] developed the first systematic non-supersingular (ordinary) curve, commonly called the MNT curve. The embedding degree k $\in \{2, 3, 4, 6\}$ is used to define MNT curves on primes.

Other methods for constructing simple curves with variable embedding degrees but higher $\rho = 2$ are proposed in schemes [11], [12].

Table 1. Curve parameter size (in bits) with associated embedding degree for obtaining the prescribed level of security [1].

| Security level (in bits) | Subgroup size $r$ (in bits) | Extension field size $q^k$ (in bits) | Embedding degree $k$ | |
|---|---|---|---|---|
| | | | $\rho = 1$ | $\rho = 2$ |
| 80 | 160 | 960-1280 | 6-8 | 3-4 |
| 112 | 224 | 2200-3600 | 10-16 | 5-8 |
| 128 | 256 | 3000-5000 | 12-20 | 6-10 |
| 192 | 384 | 8000-10000 | 20-26 | 10-13 |
| 256 | 512 | 14000-18000 | 28-36 | 14-18 |

Cocks- Pinch [13] generalised the construction of an elliptic curve of any $k$ such that $q \approx r^2$, where $q$ is the size of the area and $r$ is the prime order of the subgroup, but their method is computationally inefficient. Barreto et al. [11] explored groups in ordinary curves and introduced the BLS curves that exclude extraneous elements during Tate pairing. Under specific conditions, Brezing and Weng [14] expanded the Cocks-Pinch approach [13] and identified a polynomial family of elliptic curves with smaller $\rho$; attaining $\rho \approx 1$ remained a challenge. Kachisa et al. [15] suggested a new approach for constructing the Brezing-Weng curve [14], which has the lowest polynomials of integers in the cyclotomic field and generated elliptic curve families for $k \in \{16, 18, 36, 40\}$. Tanaka et al. [16] uncovered more patterns of families with some adaptation. The Brazing–Weng approach has also been extended by Drylo [17] to generate complete families with variable discriminant and sparse families. Barreto et al. [8] proposed a straightforward implementation approach for a prime-order elliptic curve (BN curve) with degree $k = 12$, which is the most effective and efficient elliptic curve in practice.

Scott et al. [18] extended the general MNT construction to generate some interesting curves and demonstrated their application in real-world situations. Freeman [19] proposed that existing methods of constructing prime-order elliptic curves with fixed embedding degrees could be described as a general framework. He used it to generate a curve with embedding degree 10. Scott et al. [20] recently introduced a modified family of pairing-friendly curves with embedding degree $k = 54$, which can be used to build high-security pairing-based cryptosystems. They also tried to find elliptic curves with $k = 72$ but failed due to inefficiency. Consequently, uncovering those elliptic curves into families of curves that match the constraints with the optimal embedding degree $k$ is the problem. Table 1 illustrates the parameter size of the distinct curve for selecting efficient pairing [1].

The security of a pairing-based system is defined by the difficulty of solving DLP in the finite field $F_{q^k}$ with a small value of $k$. The DLP on curve $E(F_q)$ with a subgroup of order $r$ is solved by Pollard's rho algorithm, which has a complexity of $O(\sqrt{r})$ [21]. Recent advances in state-of-the-art NFS algorithms, such as tower number field sieve (TNFS) [22], [23] have had an impact on the complexity of DLP in the extension field $F_{q^k}$. In addition, considerable progress has been observed in the special tower number field sieve (STNFS) which can optimise the DLP complexity in the extension filed $F_{q^k}$ for the composite embedding degree. Using the Brazing–Weng method, Fotiadis et al. [24] constructed families of coupling-friendly elliptic curves and derived a list of curves with optimal embedding degrees at a 128-bit level of security. To assess the security of curves, they used the STNFS complexity. Kim et al. [23] presented a new NFS method, the extended tower number field sieve (exTNFS), simplifying DLP on a finite field. ExTNFS reduced the security level of elliptic curves for pairing asymptotically. Later, Menezes et al. [25], Barbulescu et al. [26], Guillevic et al. [27] and [28] revisit the STNFS to obtain the precise finite field size for a particular protection level. For the 128-bit security level, Fotiadis et al. [29] discovered pairing-friendly curves with mixed embedding degrees. As a result, the suggestions in Table 1 are suitable for determining the curve parameters for prime $k$, but the values for overall $k$ should be modified.

Freeman, Scott, and Teske [1] released a "Taxonomy of Pairing-Friendly Elliptic Curves" in 2010, in which they documented and organised all known methods (at that time) for constructing "pairing-friendly" elliptic curves, as well as showing evidence and recommendations. In this paper, we propose to extend the Taxonomy to incorporate advances over the past decade and to update curve recommendations based on the latest attacks on the DLP. We focused on well-known pairing-friendly curves and recent attacks on attack-resistant pairing-friendly elliptic curves. We review a significant amount of literature to find a variety of strategies for constructing simple elliptic curves with specific embedding degrees and classify them using the CM discriminant value. After that, we updated Freeman et al. [1] Taxonomy by adding two new families: complete with variable discriminants (CVD) [17] and new sparse families [30] that were not covered by previous classifications [17], [20].

We demonstrate the generalisation of pairing-friendly curves and give a complete framework for generating individual and parametric families of curves with varying embedding degrees. After that, we estimate safety by considering the principal size of these families to visualise better curve families compared to previous families such as BN, KSS and BLS. We also looked at recent attacks on pairing, such as TNFS, exTNFS, and SexTNFS, and suggest that such attacks are required to increase the key size of the most popular pairing-friendly curves, such as BN, BLS, and KSS. We analyse and compare the key size expected for families of curves, while these attacks require updating the key size for pairing. As a result, we choose the potential elliptic curve from the accessible curve family. Furthermore, we recognised the lists of potential curves against STNFS attacks for a 128-bit security level that have never been discussed before, except Guillevic et al. [28].

The rest of the article is organised as follows. In section 2, we recall some basic definitions, including the elliptic curve, pairing-friendly curves, parameterisation, families of curves and related terminologies. In section 3, we present the classification of pairing-friendly curves. Some parameterised examples of pairing-friendly elliptic curves are given in Section 4. Section 5 offers the recent attacks, such as NFS, TNFS, and SexTNFS, on pairing-friendly curves. In Section 6, we explore those pairing-friendly curves that are secure against these attacks. We give discussion and recommendations in section 7 and conclude in Section 8.

## 2. Preliminaries

### 2.1. Embedding degree and $\rho$-value

**Definition 1**. *(Embedding degree). Suppose E be an elliptic curve defined over a finite field $F_q$, and let r be a prime dividing $\#E(F_q)$. The embedding degree of E with respect to r is the smallest k such that $q^k - 1$, but does not divide $q^i - 1$ for all $0 < i < k$.*

The embedding degree $k$ determines the feasibility of pairing construction in terms of security and efficiency. Curves with a large $k$ cannot be regarded as acceptable for pairing implementation. The only curves that can be paired are those with a minimal embedding degree. Consider an elliptic curve $E(F_q)$ over a finite field $F_{q^2}$, with subgroup $log(r) = 160$ bits and $log(q) = 512$ bits. There is a transformation from the elliptic curve $E(F_q)$ to $F_{q^2}$, where $q^2 = 1024$ bits. The embedding degree is represented by exponent 2. According to Barreto et al. [8], the elliptic curve with embedding degree $k = 12$ is ideal.

Recall that the hardness of the DLP on the elliptic curve $E(F_q)$ of order $r$ and the finite field extension $F_{q^k}$ of order $q$ determine the security of coupling-based cryptographic systems. Both problems are defined in their respective group order, $r$ and $q$. As a result, we must define $\rho = \log(q) / \log(r)$, which evaluates the correlation of order $r$ and $q$. For $\rho \geq 1$, the dense coordinates of the points are produced by elliptic curves due to the vast expansive finite field $F_{q^k}$, which affects the arithmetic efficiency, resulting in $k\rho = \log(q^k) / \log(r)$ happens. Suppose $q(x), r(x), t(x) \in Q[x]$ and $(q, r, t)$ parameterises a family of curves with embedding degree k. The ρ-value of $(q, r, t)$ is represented as $\rho(q, r, t) = \deg(q) / \deg(r)$. Elliptic curves with small $\rho$-values are considered robust and computationally efficient. For instance, a curve of 512-bits size of subgroup with $\rho = 1$ is built over an extension field size 512-bits, while the same curve with $\rho = 2$ is built over an extension field size 1024-bits. The group operations are more efficient in the first case.

### 2.2. Pairing-friendly curve

**Definition 2**. *(Elliptic curve). An elliptic curve E over a field K, denoted as E/K (Char(K) ≠ 2,3) is the set of points in K × K of an equation $y^2 = x^3 + ax + b$, where $a, b \in K$ and $\Delta = -16(4a^3 + 27b^2) \neq 0$, together with some point O at infinity.*

Here, K can be a finite field $F_q$ for a prime $q > 3$ or a field extension $F_{q^k}$ such that the set

$$E(F_q) = \{(x, y) \in (F_q, F_q) | y^2 = x^3 + ax + b\} \cup \{O\} \tag{1}$$

is known as the group of $F_q$-rational points of E. The group $E(F_q)$ is known as the abelian group, in which O denotes the point at infinity. The number of points, also known as the order of group $E(F_q)$ is defined as the $\#E(F_q) = r = q + 1 - t$, (Hasse's Theorem) such that $|t| \leq 2\sqrt{q}$, where $t$ is the trace of Frobenius.

**Definition 3**. *(r-torsion point). Suppose E be an elliptic curve defined over $F_q$, and $E(F_q)$ denotes the group of $F_q$-rational points. For some integer $r$, we suppose $E[r]$ be the group of r-torsion and $E(F_q)[r]$ the group of r-torsion points of E defined over $F_q$. The set of r-torsion points in $E(F_q)$ is defined as*

$$E(F_q)[r] = \{P \in E(F_q) | [r]P = 0\} \tag{2}$$

Suppose $G_1$ and $G_2$ are two subgroups of $E(F_q)$ and $G_T$ be a subgroup of $F_{q^k}$, both of order r; there is an asymmetric pairing $e: G_1 \times G_2 \to G_T$. The suitable combination of $q, k, E$ and $r$ constitute a parameter for constructing a pairing-based cryptographic protocol.

**Definition 4.** *(Pairing-friendly elliptic curve). An elliptic curve $E(F_q)$ is said to be pairing-friendly if the following conditions are satisfied:*
- *there exists a prime $r \geq \sqrt{q}$ such that $r | E(F_q)$,*
- *the value $\rho = \frac{\log(q)}{\log(r)} \leq 2$, close to 1, while the embedding k with respect to r satisfies $k \leq \log(r)/8$.*
- *Value r must be large enough so that DLP in $G_1$ and $G_2$ is computationally hard.*
- *Embedding degree k is large enough so that DLP in $F_{q^k}$ means $G_T$ is equally hard as in $G_1$ and $G_2$, but is small enough for efficient operations in $G_T$.*

### 2.3. Parameterisation of pairing-friendly elliptic curve

Now, we discuss the strategy to parameterise the traces of curves [8]. We fix a polynomial $t(x)$, a trace of Frobenius, and construct the polynomials $r(x)$ and $q(x)$. More concretely, the CM algorithm can be used to construct an elliptic curve over $F_q(x_0)$ with $r(x_0)$ points with embedding degree k, if $q(x_0)$ is prime for some $x_0$.

**Theorem 5.** [19] *Choose an integer $k > 0$, and suppose $\Phi_k(x)$ be the $k^{th}$ cyclotomic polynomial. Let $t(x)$ be a polynomial of trace with integer coefficient, $r(x)$ be an irreducible factor of $\Phi_k(t(x) - 1)$, and let $f(x) = 4q(x) - t(x)^2$. Set a positive square-free integer to be D and let there exists a solution for equation $Dy^2 = f(x)$ be $(x_0, y_0)$ in which $q(x_0)$ and $r(x_0)$ are prime.*

*For a small D, there exists an efficient CM algorithm that constructs an elliptic curve E defined over $F_q(x_0)$ such that $E(F_q(x_0))$ has prime order $r(x_0)$ and E has embedding degree at most k.*

*Proof.* The well-known solution to the equation $Dy^2 = f(x)$ is $(x_0, y_0)$, in which $q(x_0)$ is prime. For a small D, the CM algorithm constructs an elliptic curve $E$ over $F_q(x_0)$ of order $\#E(F_q(x_0)) = r(x_0)$. Elliptic curve $E(F_q(x_0))$ has prime order since $r(x_0)$ is prime. Lemma 1 of Barreto et al. [31] proves that E with embedding degree k has prime order $r(x_0)$ that divides $\Phi_k(t(x_0) - 1)$ but not $\Phi_l(t(x_0) - 1)$ for $l < k$. Because we selected the polynomial $r(x)$ to divide $\Phi_k(t(x) - 1)$, so it is guaranteed that $r(x_0)$ divided $q(x_0)^k - 1$ and the embedding degree of E is at most k.

### 2.4. Families of pairing-friendly curves

To construct a curve of prime order with embedding degree $k$, we must first choose the polynomials $q(x), r(x)$ and $t(x)$ that fulfill the conditions of Theorem 5 on various values of $x$ until $r(x)$ and $q(x)$ are prime.

**Definition 6.** Suppose $q(x), r(x)$, and $t(x)$ be nonzero polynomials with integer coefficients. For any given positive integer $k$ and positive square-free integer $D$, we say that polynomial triplet $(q(x), r(x), t(x))$ parameterises a family of the pairing-friendly ordinary curve if the following conditions are fulfilled.

1. $q(x) = p(x)^d$ for some $d \geq 1$ and $p(x)$ is prime.
2. $r(x) = c.r'(x)$ with $c \geq 1 \in Z$, and $r'(x)$ is prime.

3. q(x) + 1 − t(x) = h(x)r(x) for some h(x) ∈ Q.
4. r(x)|($\Phi_k$t(x) − 1), where $\Phi_k$ is the k-the cyclotomic polynomial.
5. The CM equation $Dy^2 = 4q(x) − t(x)^2$ has infinitely many solutions.

In this way, we can define a family of curves and construct an elliptic curve with embedding degree k and the CM discriminant D. In practice, and it is easy to find $r(x)$ and $q(x)$ that fulfilled the four conditions of definition 6. However, choosing these parameters is quite difficult because $Dy^2 = 4q(x) − t(x)^2$ responds to infinitely many solutions. Usually, if $f(x)$ is a square-free polynomial with a degree of at least 3, then a finite number of solutions exist to equation $Dy^2 = f(x)$. Therefore, we ensure that parameters $(q, r, t)$ can denote a family of curves.

**Corollary 7**. *Choose an integer $k > 0$ and polynomials $q(x)$, $r(x)$ and $t(x) \in Z[x]$ that satisfy the first three conditions of definition 6. Let $f(x) = 4q(x) − t(x)^2$. Consider $f(x) = ax^2 + bx + c$, where $a, b, c \in Z$, such that $a > 0$ and $b^2 − 4ac \neq 0$. Suppose a square-free integer is D such that aD is not a square. If there exists a solution $(x_0, y_0)$ of equation $Dy^2 = f(x)$, then tuple $(q, r, t)$ denotes a family of curves with embedding degree k.*

It is clear from Theorem 5 that if $f(x)$ is square-free and quadratic, we can obtain the family of curves of the prescribed degree for each $D$. On the other side, if $f(x)$ is a linear function time a square, we can still get the family of curves, but for only a single $D$. Barreto and Naerhig utilise the identical method to construct the curve with embedding degree $k = 12$.

## 3. Revised Taxonomy of Pairing-friendly elliptic curve

In 2010, Freeman, Scott, and Teske published a "Taxonomy of Pairing-Friendly Elliptic Curves" in which they catalogue and organise all the known methods (at that time) for generating "pairing-friendly" elliptic curves and give examples and recommendations. The pairing-friendly curve construction methods are classified into individual and parametric families of curves. The methods for constructing individual (not in a family) curves consider two parameters: $q$ and $r$; that is, there exists an elliptic curve E over finite field $F_q$ with subgroup $r$ and embedding degree $k$. While the methods for constructing the family of curves assume the polynomial $q(x)$ and $r(x)$ in such a way that if $q(x_0)$ is prime power for some $x_0$, there must be an elliptic curve E over $F_{q(x_0)}$ with subgroup $r(x_0)$ and embedding degree $k$. Fig. 1 demonstrates the revised classification of a pairing-friendly elliptic curve. Here, we extend the taxonomy to incorporate advances over the past decade and to update curve recommendations based on the latest attacks on DLP.

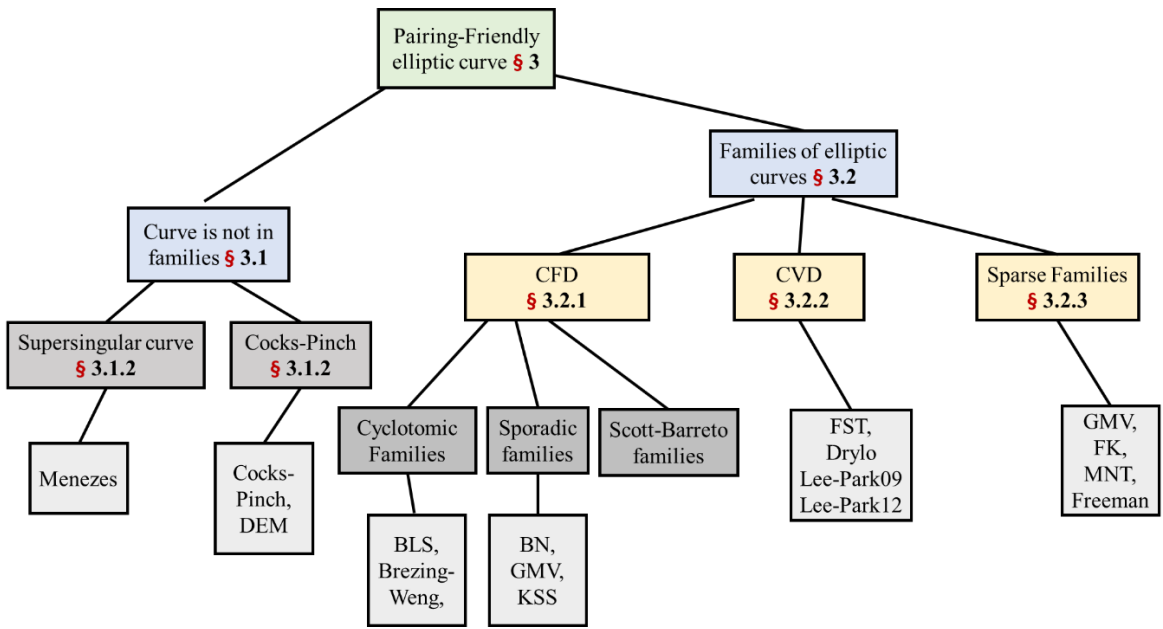

Fig. 1. Revised Taxonomy of pairing-friendly elliptic curve

## 3.1. Elliptic curve not in the family

*Supersingular Curve.* An elliptic curve is said to be a supersingular curve if and only if $t = 0 \pmod{q}$ and the embedding degree k ∈ {1,2,3,4,6} otherwise ordinary curve. Supersingular elliptic curves with embedding degrees 4 and 6 are defined at characteristics two and three, respectively. Menezes et al. [9] discussed prime-order supersingular curves with embedding degree k ∈ {3,4,6}. These curves are considered weak because of the existence of efficient pairs defined on these curves, so they were not suitable for building cryptographic systems. Some researchers do not like supersingular curves because $k = 2$ and 6 are too small, so they focused on simple curves of prime order $n$, such as $r = n$.

*Ordinary Curve.* Since the supersingular curve is limited to small embedding degrees, for prime fields $k = 2$ and $k \leq 6$ in general, we move on to ordinary curves. Cocks-Pinch [13] and Dupont et al.'s [12] methods produce simple curves with small embedding degrees; however, they do not fall in families. The main problem in ordinary curves is determining the pairing-friendly elliptic curve parameter $(q, r, t)$ that satisfies the conditions in Definition 6. In both methods, the trace of Frobenius $t$ is increased by some integer in $Z_r$. Thus, the value of t is approximately the same as the size of $r$, which means that the $\rho$-value is approximately 2. In practice, such a choice of parameters is inefficient for pairing computations for the most well-known Tet pairing variants, such as Ate and Twisted pairing. Algorithm 1 defines the Cocks-Pinch method for constructing simple curves of composite order.

---
**Algorithm 1. Cocks-Pinch construction of composite order** [13]

*To fix k and D, do the following steps:*
1. Choose a prime $r$ such that $k|(r-1)$ and $\frac{(-D)}{2} = 1$
2. Evaluate z, a primitive k-th root of unity in $Z_r$ and $t_0 = z + 1$
3. Evaluate $t_0 = \frac{t_0 - 2}{\sqrt{-D}} \mod r$
4. Reduce $t = t_0 \mod r$ and $y = y_0 \mod r$, let $q = (t^2 + Dy^2)/4$

---

## 3.2. Families of pairing-friendly curves

The basis of generating pairing-friendly elliptic curves is to find parameters $(q, r, t)$ that satisfy the constraints. The Cocks–Pinch approach [10] and the Dupont–Ange–Morran (DEM) method [5] are the two main methods for identifying such triples as polynomials family $(q(x), r(x), t(x)) \in Q[x]$. Based on the complex multiplication (CM) polynomial $f(x) = 4q(x) - t(x)^2$, elliptic curves are categorised into three families: complete families with fix discriminant (CFD) [8], [14], [15], [31], complete families with variable discriminant (CVD) [1], [17], [32], and sparse families [10], [19], [30], shown in Fig. 1. Their constructions are depending on the ability to find integers $x$, and y satisfying the CM equation. Table 2 summarises the category of the pairing-friendly elliptic curve family.

Table 2. Families of the pairing-friendly curve

| Family | CM |
|---|---|
| CFD | $f(x) = Dy(x)^2$, for square-free $D > 0$, |
| CVD | $f(x) = g(x)y(x)^2$, for some $g(x) \in Q[x]$ with deg g=1 |
| Sparse | $f(x) = g(x)y(x)^2$, for some $g(x)$ is quadratic not perfect square |

### 3.2.1 Complete families fix discriminant

To construct the CFD family of curves, we find the polynomials $q(x), r(x), t(x)$ that fulfil certain divisibility conditions, and the equation has infinitely many solutions $(x, y)$. The complete family construction chooses the parameters $D, t(x), r(x), p(x)$ in such a way that $4h(x)r(x) - (t(x) - 2)^2$ is D times the square of polynomials $y(x)$. Scott et al. [18] and Barreto et al. [31] are the two methods for constructing complete families, which is generalised from Brezing et al. [14] method.

> *Algorithm 2. Brezing-Weng Construction for the cyclotomic family of curves* [14]
>
> *On given fixed integer k and positive square-free integer D, performs the following steps:*
> 1. *Picks a number field K containing $\sqrt{-D}$ and primitive $k^{th}$ root of unity $\zeta_k$.*
> 2. *Compute an irreducible polynomial $r(x) \in Z[x]$ such that $Q[x]/r(x) \cong K$*
> 3. *Suppose a polynomial be $t(x) \in Q[x]$ that map to $\zeta_k + 1 \in K$.*
> 4. *Suppose a polynomial be $y(x) \in Q[x]$ that map to $(\zeta_k - 1)/\sqrt{-D} \in K$.*
> 5. *Suppose a $q(x) \in Q[x]$ be defined as $(t(x)^2 + Dy(x)^2)/4$.*
> 6. *If the parameters $p(x)$ and $r(x)$ are prime, then the triple $(q(x), r(x), t(x))$ denotes the family of the curve with embedding k and discriminant $\sqrt{-D}$.*

***Cyclotomic families of curves.*** Barreto et al. [31] and Brezing et al. [14] extended the Cocks-Pinch method by parameterising $(q, r, t)$ as polynomials and reduced the ρ-value. Brezing et al. [14] discussed the fullest generality of the construction. Barreto et al. [31] presented the first construction of a cyclotomic family of curves by considering the polynomial $r(x)$ defining number field K to be kth cyclotomic polynomial. It selects $\zeta_k \in K$ and uses the constraints that if k is divisible by 3, then $\sqrt{-3} \in K$. Brezing et al. [14] discussed more general construction by fixing $r(x)$ to be cyclotomic polynomial $\Phi_l t(x)$ of a few multiple $l$ of the desire embedding degree $k$ and selecting distinct representative of $\zeta_k \in Q[x]/(r(x))$. In both constructions, the discriminants D are usually 1 or 3, and cyclotomic polynomials must satisfy condition 2 of definition 6. *Algorithm 2* shows that Brezing-Weng construction for a cyclotomic family of curves

***Sporadic families of curves.*** In order to construct the elliptic curve for the cyclotomic family, Brezing and Weng only assume cyclotomic polynomials r(x). However, the extension of cyclotomic fields, defined by the non-cyclotomic polynomials, yields more effective results. In order to construct such an extension field, it is to examine the cyclotomic polynomials $\Phi_l(x)$ on any polynomial u(x). We have some advantages for constructing the curve for any factorisation of cyclotomic polynomial $\Phi_l(u(x))$. Otherwise, as is usually the case, it is required to evaluate the parameter $(q, r, t)$ at $u(x)$. Galbraith et al. [33] examined that $\Phi_l(u(x))$ is factorised if u is quadratic and $\Phi_l$ has degree 8. For $l = 12$, there are two $u(x)$ that factorised the $\Phi_l(x)$. BN curve constructed by Barreto et al. [8] uses one such factorisation.

***Scott-Barreto families of curves.*** The Scotte-Barreto family of curve assume K to be an extension of the cyclotomic field, but it does not contain an element $\sqrt{-D}$. For any polynomial $t(x)$ and $r(x)$ to be irreducible to $\Phi_k(t(x) - 1)$, the $Q[x]/r(x)$ evaluate an extension of a cyclotomic field.

*3.2.2    Complete with variable discriminant*

In the CFD family, the methods given by Brezing and Weng assume that the CM discriminant D is known in advance and based on such D all curves are constructed. For instance, the mostly curves constructed by Barreto, Lynn and Scott, and Brezing and Weng demand $D = 3$. On another side, there exist some families of a pairing-friendly elliptic curve with variable CM discriminant D. For higher security it is required to utilise the curves with sufficiently large discriminant. Thus, Freeman et al. [1] addressed complete families with variable discriminant.

**Definition 8**. *A tuple of polynomials $(q(x), r(x), t(x))$ parameterises a complete family of elliptic curves with embedding degree k and variable discriminant if it fulfils the first four conditions of Definition 6, and $4q(x) - t(x)^2 = xh(x)^2$ for some $h(x) \in Q[x]$. Now, substitute $x = Dx^2$, such that $D > 0$ is a square-free integer, outputs a complete family $q(Dx^2), r(Dx^2), t(Dx^2)$ with discriminant D that satisfied the first two conditions of Definition 6.*

Freeman et al. [1] got the variable discriminant families from complete families $(q, r, t)$ with discriminant D in such a way that there must be the polynomials $q', r', t', y' \in Q[x]$ such that $q(x) = q'(x^2), r(x) = r'(x^2), t(x) = t'(x^2)$ and $4q(x) - t(x)^2 = Dx^2(y'(x^2))^2$. These family $(q', r', t')$ satisfies definition 8 on $x \leftarrow \frac{x}{D}$. Drylo et al. [17] illustrated a construction for a complete family with variable discriminant, as shown in *Algorithm 3*.

> **Algorithm 3. Drylo Construction for the complete family with variable discriminant** [17]
> On the given number field $K$ containing kth root of unity $\zeta_k$, performs the following steps:
> 1. Picks $z \in K$ in such a way that $a = -z^2$ is a primitive element of $K$.
> 2. Suppose a polynomial $r(x)$ is $a$, and set $K = Q[x]/r(x)$.
> 3. Suppose a polynomial be $t(x) \in Q[x]$ that map to $\zeta_k + 1 \in K$.
> 4. Suppose a polynomial be $h(x) \in Q[x]$ that map to $(\zeta_k - 1)/\sqrt{-x} \in K$.
> 5. Suppose a $q(x) \in Q[x]$ be defined as $(t(x)^2 + Dy(x)^2)/4$.
> 6. If the parameters $q(x)$ and $r(x)$ are prime, then the triple $(q(x), r(x), t(x))$ denotes the family of the curve with embedding $k$ and variable discriminant.

### 3.2.3 Sparse families of elliptic curve

The primary idea of a sparse family of the curve is based on the Miyaji et al. [10] curve. Most of the sparse family curves are prime order, but they are restricted to embedding degree $k \leq 10$. To construct a sparse family of curves, we choose the polynomials $q(x)$, $r(x)$ and $t(x)$, which satisfies the first four conditions of Definition 6 and for which the CM equation $Dy^2 = 4q(x) - t(x)^2 = 4h(x)r(x) - (t(x) - 2)^2$ has many infinite solutions, where $h(x)$ is the cofactor, satisfying $h(x)r(x) = q(x) - 1 + t(x)$. For $h(x) = 1$, we can obtain a curve of prime order. Miyaji et al. [10] were the first to construct the ordinary elliptic curves of prime order with a prescribed embedding degree. They realise that if $4h(x)r(x) - (t(x) - 2)^2$ is a quadratic polynomial, we can transform the equation into a generalised Pell equation. Thus, it gives an infinite number of solutions to get a family of curves. *Algorithm 4* presents the construction of the MNT curve with embedding degree $k$.

> **Algorithm 4 Construction of MNT curve with embedding degree $k$** [10]
> **Input**: Parameters $q(u)$, $t(u)$ and $k$.
> **Output**: Set of points for MNT curve
> 1. Take the parameters, $q(u)$ and $t(u)$ associated with the elliptic curve with embedding degree $k$. Using the Hass theorem, compute $r(u)$ as $r(u) = q(u) + 1 - t(u)$.
> 2. Pick an integer for $u$ until we find $u_0$ such that $p(u_0)$ and $r(u_0)$ are both primes.
> 3. Using CM algorithm, find the solutions to the norm equation $t^2 - 4q = DV^2$, such that $|D|$ is small. Repeat this step until we have a small $|D|$.
> 4. Find the solutions to the norm equation $t^2 - 4q = DV^2$.
> 5. If $d = -3D$ is positive and square-free, it can have infinitely many solutions.

Freeman et al. [19] exhibit such a curve by considering that if $f(x) = 4q(x) - t(x)^2$ is square free, the equation defines the affine plane of genus $g = \lfloor \frac{\deg f - 1}{2} \rfloor$. If $f(x)$ is quadratic, then genus is zero that has no solutions or infinite many integral solutions. We obtain a family $(q, r, t)$ for infinite solutions using Definition 6.

## 4. Some Parameterise use of Pairing-friendly elliptic curve

To select an elliptic curve for pairing construction, we must parameterise the number of points on the curve and the size of the field defined by the polynomials as $r(x)$ and $q(x)$, respectively. For each $x_0$, CM algorithm constructs an elliptic curve on given prime values for polynomials $r(x_0)$ and $q(x_0)$. Some well-known elliptic curves are given by Miyaji et al. [10], Barreto et al. [31], Barreto et al. [8], Cocks et al. [13], Brezing et al. [14], Kachisa et al. [15], and Freeman [19].

### 4.1. Individual elliptic curve

**Cocks-Pinch Method.** The Cocks and Pinch provided the general construction for curves of arbitrary embedding degree $k$. Algorithm 1 discusses the construction of an elliptic curve over $F_q$ of embedding degree $k$. The drawback of this method is the large value of $\rho$. The size of the field $F_q$ is related to the subgroup of prime $r$ in such a way that $q = r^2$ yields the value of $\rho = 2$. Thus, Cocks-Pinch method leads to inefficient implementation and so impractical.

## 4.2. Complete family with fixed discriminant

Brezing and Weng presented the state-of-art method for complete family construction, which starts by selecting a fixed embedding degree and square-free CM discriminant D [14]. The complete family with a small discriminant D is fascinating for implementation. However, such families are easier to attack by various attacks on DLP.

***Brezing-Weng method.*** Brezing and Weng [14] adapted the generalisation of Cock and Pinch [13] method and constructed the families of pairing-friendly elliptic curves with small ρ-values. This method has a similar form but instead of using polynomial $r(x)$, it makes use of $r$ and generates the PF-EC having ρ-values closer to 1. However, finding a suitable polynomial $r(x)$ is challenging. For $k < 1000$, and k is odd, the parameters $(q(x), r(x), t(x))$ for complete family of a pairing-friendly elliptic curve with embedding degree k and discriminant 1 are as:

$$q(x) = \frac{1}{4}(x^{2k+4} + 2x^{2k+2} + x^{2k} + x^4 - 2x^2 + 1)$$
$$r(x) = \Phi_{4k}(x)$$
$$t(x) = -x^2 + 1$$

This family has ρ-value $\frac{k+2}{\phi(k)}$.

***BLS Curve with embedding degree*** $k \in \{12, 48\}$. In 2002, Barretto, Lynn and Scott [31] reported well-known elliptic curves, known as BLS curves, that are suitable for constructing optimal ate pairing. BLS curves have a small discriminant parameter that yields a simple equation from which we can derive PF-EC of embedding degree $k = 0 \ (mod \ 6)$ with a maximal twist of $d = 6$, and have a relatively small ρ value, as defined as $\rho = \frac{2+\frac{k}{3}}{\phi(k)}$.

The BLS curve is a special elliptic curve over a finite field $F_q$ defined by the equation of form $E: y^2 = x^3 + b$. Like BN curve, it has a twist of order 6 but does not have a prime order. Whilst its order is divisible by a large parameterised prime $r$, the pairing is defined on $r$-torsion points. On distinct embedding degrees, BLS curves vary. Here, we discuss the BLS12 and BLS48 curves families with embedding degrees 12 and 48, respectively. Table 3 defines the parameters $q$ and $r$ to construct the BLS curves with embedding degrees 12 and 48.

Table 3 Complete family of the curve with embedding degrees 12 and 48, using BLS method

| | | |
|---|---|---|
| **BLS12:** | $q = \frac{(u-1)^2(u^4-u^2+1)}{3} + u,$ | $r = u^4 - u^2 + 1$ |
| **BLS48:** | $q = \frac{(u-1)^2(u^{16}-u^8+1)}{3} + u,$ | $r = u^{16} - u^8 + 1$ |

Where, $u$ is a random chosen integer.

***BN curve with embedding degree k=12.*** In 2005, Barrato and Naehrig [8] presented a PF-EC, generally known as the BN curve, which is considered a cryptography breakthrough. An Ate pairing construct over BN curves is considered the optimal pairing. An elliptic curve $E$ over finite field $F_q$, where $q \geq 5$, such that $q$ and curve $E$ order $r$ are prime numbers and defined by

$$q = 36u^4 + 36u^3 + 24u^2 + 6u + 1$$
$$r = 36u^4 + 36u^3 + 18u^2 + 6u + 1$$

where, $u$ is random chosen integer. BN curves are suitable for pairing with a strong level of security and achieve efficient pairing-based cryptographic primitives. For example, a pairing over a 256-bit BN curve is considered the secure method for signing a message with a small signature size.

***KSS curve with embedding degree*** $k \in \{8, 16, 18, 32, 36, 40\}$. Kachisa et al. [15] put forward the idea of Brezing et al. [14] and discussed a new method to construct the PFEC. This method uses a minimal polynomial of elements other than cyclotomic polynomial $\Phi_l(x)$ to define the of the cyclotomic field. This method is constructing a new family of PF-EC with embedding degree $k \in \{8,16,18,32,36,40\}$. Table 4 summarises the polynomial of parameters $(q(u), r(u), t(u))$ for complete families of curves with embedding degree $k \in \{8,16,18,32,36,40\}$ and distinct $\rho$ −values by KSS method.

Table 4 Complete families of the curve with embedding degree $k \in \{8,16,18,32,36,40\}$ by KSS method

| |
|---|
| **KSS: $k = 8, D = 3, \rho = \frac{5}{4}$** |
| $q(u) = \frac{1}{3}(u^{10} + u^9 + u^8 - u^6 + 2u^5 - u^4 + u^2 - 32u + 1)$ <br> $r(u) = u^8 - u^4 + 1, \qquad t(u) = u^5 - u + 1$ |
| **KSS: $k = 8, D = 1, \rho = \frac{3}{2}$** |
| $q(u) = \frac{1}{180}(u^6 + 2u^5 - 3u^4 + 8u^3 - 15u^2 - 8u + 125)$ <br> $r(u) = u^4 - 8u^2 + 25, \qquad t(u) = \frac{1}{15}(2u^3 - 11u + 15)$ |
| **KSS: $k = 16, D = 1, \rho = \frac{5}{4}$** |
| $q(u) = \frac{1}{980}(u^{10} + 2u^9 + 5u^8 + 48u^6 + 152u^5 + 240u^4 + 625u^2 + 2398u + 3125)$ <br> $r(u) = u^8 + 48u^4 + 625, \qquad t(u) = \frac{1}{35}(2u^5 + 41u + 35)$ |
| **KSS: $k = 18, D = 3, \rho = \frac{4}{3}$** |
| $q(u) = \frac{1}{21}(u^{10} + 5u^7 + 7u^6 + 37u^5 + 188u^4 + 259u^3 + 343u^2 + 1763u + 2401)$ <br> $r(u) = u^6 + 37u^3 + 343, \qquad t(u) = \frac{1}{7}(u^4 + 16u + 7)$ |
| **KSS: $k = 32, D = 1, \rho = \frac{9}{8}$** |
| $q(u) = \frac{1}{2970292}(u^{18} - 6u^{17} + 13u^{16} + 57120u^{10} + 344632u^9 + 742560u^8 + 815730721u^2 \\ \qquad -4948305594u + 1060449373)$ <br> $r(u) = u^{16} + 57120u^8 + 815730721, \qquad t(u) = \frac{1}{3107}(-2u^9 + 56403u + 3107)$ |
| **KSS: $k = 32, D = 3, \rho = \frac{7}{6}$** |
| $q(u) = \frac{1}{28749}(u^{14} + 46u^{13} + 7u^{12} + 683u^8 - 2510u^7 + 4781u^6 + 117649u^2 - 386569u + 823543)$ <br> $r(u) = u^{12} + 683u^6 + 117649, \qquad t(u) = \frac{1}{259}(2u^7 + 757u + 259)$ |
| **KSS: $k = 40, D = 1, \rho = \frac{11}{8}$** |
| $q(u) = \frac{1}{1123380}(u^{22} - 2u^{21} + 5u^{20} + 6232u^{12} + 10568u^{11} + 31160u^{10} + 9765625u^2 - 13398638u \\ \qquad + 48828125)$ <br> $r(u) = u^{16} + 8u^{14} + 39u^{12} + 112u^{10} - 79u^8 + 2800u^6 + 24375u^4 + 125000u^2 + 390625,$ <br> $t(u) = \frac{1}{1185}(2u^{11} + 6469u + 1185)$ |

***Scott-Guillevic with embedding degree k=54***. In practice, the elliptic curve with embedding degree 48 suggested by Barreto et al. [31] achieved an AES-256 level of security. The curves with embedding degree up to 50 is only considered by freeman et al. [1]. Recently, Scott et al. [20] noticed a new curve using KSS method, but the curve with embedding degree 54 is not the kind of KSS curve. It observed that such a family of the curve was occasional and unrelated to any existing family. However, it exhibits a certain pattern, which shows some possibility that it might be a member of an undiscovered family of families. Thus, the authors announced the new family of pairing-friendly elliptic curves with embedding degree 54, which makes the DLP's difficulty more strong since it is used for pairing-based cryptography. There exists an element $-\zeta_{54} - \zeta_{54}^{10} \in \mathbb{Q}(\zeta_{54})$ that, using the KSS method, yields the following parameters

$q(u) = 310u^{20} + 310u^{19} + 39u^{18} + 36u^{11} + 36u^{10} + 35u^9 + 3u^2 + 3u + 1$
$r(u) = 39u^{18} + 35u^9 + 1$
$t(u) = 35u^{10} + 1$
$c(u) = 3u^2 + 3u + 1$

Where c is a cofactor. The total number of points on the curve is $\#E = cr$. Since $4q - t^2 = 3f^2$ for some polynomial $f$, it can be observed that the CM discriminant is 3. Therefore, like BN and BLS curves, it has twists of degree 6 that provides good optimisation.

## 4.3. Complete families with variable discriminant

Recall family of curves with large discriminant values are less susceptible to various attacks on DLP. In this respect, complete families with variable discriminant use CM discriminant of polynomial representation. In this family, the CM polynomial has $f(x) = g(x)y(x)^2$ and $\deg(g) = 1$. Such families can be constructed using the Brezing-Weng method by substituting the square-free D with the linear term $g(x)$ in such a way that $\sqrt{-g(x)} \in Q[x]/r(x)$. This family exhibits preferable CM-discriminant but with limited choices. In practice, to evaluate suitable parameters for variable CM discriminant, we are looking for $x_0 = Dy^2 \in Z$ such that $r(x_0)$ and $p(x_0)$ are prime.

Implementing curves with prime embedding degree k has become more important as attacks on the finite fields in which pairs are defined have become more important. Nanjo et al. [34] introduced a technique to generate efficient final exponentiation algorithms for a certain cyclotomic family of curves with arbitrary prime k of k =1 (mod 6). When the proposed method is used for multiple curves with k = 7, 13, and 19, it is found that it generates an algorithm like the previously state-of-the-art lattice-based method.

**Robert Drylo method for CVD.** Drylo constructed the complete families with variable discriminants by generalising the Brezing-Weng method. This method enables us to find families with few embedding degrees, which improve on ρ -values with parameter $\rho = \frac{max\{2degt, 1+2degh\}}{degr}$, which is equal to $\rho = \frac{2degr-1}{degr}$.

Table 5 Complete families with variable discrepant with embedding degree $k \in \{8,16,18,32,36,40\}$ by Drylo

| |
|---|
| $k = 8, D = 1, 11 (mod\ 24), \rho = \frac{7}{4}$ |
| $q(u) = \frac{1}{576}(4u^7 - 39u^6 + 170u^5 - 311u^4 + 52u^3 + 716u^2 - 384u + 196)$<br>$r(u) = u^4 - 4u^3 + 8u^2 + 8u + 4, \quad t(u) = \frac{1}{12}(-u^3 + 5u^2 - 16u + 14)$ |
| $k = 9, D = 1, \rho = \frac{5}{3}$ |
| $q(u) = \frac{1}{4}(59049u^{10} + 6561u^9 + 8748u^8 + 2916u^7 + 972u^6 + 1296u^5 + 108u^4 + 36u^3 + 12u^2 + u + 1)$<br>$r(u) = 729u^5 + 27u^3 + 1, \quad t(u) = 243u^5 + 1$ |
| $k = 15, D = 1, \rho = \frac{13}{8}$ |
| $q(u) = \frac{1}{4}(531441u^{13} - 236196u^{11} + 39366u^{10} + 39366u^9 - 8748u^8 - 729u^7 + 486u^6 - 243u^5 + 135u^4 + 18u^3 + 18u^2 + u + 1)$<br>$r(u) = 6561u^8 - 2187u^7 + 243u^5 - 81u^4 + 27u^3 - 3u + 1, \quad t(u) = 9u^2 + 1$ |
| $k = 28, D = 1, 3, \rho = \frac{3}{2}$ |
| $q(u) = \frac{1}{4}(2624144u^{18} + 65536u^{17} - 32768u^{15} + 16384u^{14} + 12288u^{13} - 3072u^{11} + 2816u^9 - 192u^7 + 48u^5 + 16u^4 - 8u^3 + u + 1)$<br>$r(u) = 4096u^{12} - 1024u^{10} + 256u^8 - 64u^6 + 16u^4 - 4u^2 + 1, \quad t(u) = 512u^9 + 1$ |
| $k = 30, D = 1, \rho = \frac{13}{8}$ |
| $q(u) = \frac{1}{4}(244140625u^{13} + 195312500u^{12} + 78125000u^{11} + 19531250u^{10} + 2353750u^9 - 140625u^9 - 43750u^6 - 6875u^5 - 125u^4 + 150u^3 - 50u^2 + 9u + 1)$<br>$r(u) = 390625u^8 + 78125u^7 - 3125u^5 - 625u^4 - 125u^2 + 5u + 1, \quad t(u) = -25u^2 + 1$ |

## 4.4. Sparse Families

Identical to complete families with variable discriminant, sparse families have a large discriminant, so they avoid any attack on DLP. In this family, the CM polynomial has the form $f(x) = g(x)y(x)^2$ and deg g is 2 and non-square. However, several sparse family curves have prime order, restricted to embedding degree k ≤ 10. Such families of curves fulfil 1- 4 conditions of definitions on given polynomials $q(x), r(x)$ and $t(x)$ and help respond to many infinite solutions in CM equation $Dy^2 = 4q(x) - t(x)^2 = 4h(x)r(x) - (t(x) - 2)^2$, where $h(x)$ is the cofactor. It transforms the CM polynomial $4h(x)r(x) - (t(x) - 2)^2$ to the generalised Pell equation.

**MNT curves with embedding degree $k \in \{3, 4, 6\}$**. In 2002, Miyaji, Nakabayashi and Takano [10] described an explicit approach for constructing ordinary curves $E(F_q)$ of prime order $n = r$ with embedding degree $k \in \{3, 4, 6\}$. Suppose a prime $q$, and $E(F_q)$ be elliptic curves such that $r = \#E(F_q)$ is prime. Let $t = q + 1 - r$ be a trace. Then, the elliptic curve has embedding degree $k$ and if only if there must be $u \in \mathbb{Z}$ such that $q(u)$ and $t(u)$ are given in Table 6.

Table 6 parameterisation for MNT curves with embedded degree $k \in \{3, 4, 6\}$

| k | q(u) | t(u) |
|---|------|------|
| 3 | $12u^2 - 1$ | $-1 \pm 6u$ |
| 4 | $u^2 + u - 1$ | $-u$ or $u + 1$ |
| 6 | $4u^2 + 1$ | $1 \pm 2u$ |

For embedding degree $k \in \{3, 4, 6\}$, the CM equation $Dy^2 = 4q(x) - t(x)^2$ determines an elliptic curve of genus zero, with the right-hand side, are quadratic in $x$. Using linear changes of variables, this CM equation can be transformed into the generalised Pell equation $x^2 - SDy^2 = M$. Now, the method to find the solution to the generalised Pell equation is first to find the minimal positive integer solution. *Algorithm 2* summarises to find the solution of MNT curve.

**Scott and Barreto method**. Scott and Barreto [18] extended the MNT method by aiding a small constant cofactor h. They fix small integers h and d and replace r with $\Phi_k(t - 1)/d$ and $t$ with $x + 1$ in equation (6) which gives

$$Dy^2 = \frac{4h\Phi_k(x)}{d} - (x - 1)^2 \quad (6)$$

It can be observed that the right-hand side of Eq. (6) is quadratic in $x$ for embedding degree $k \in \{3, 4, 6\}$ as like the MNT curve. So, we write Eq. (6) into a generalised Pell equation. Now, MNT is applied to explore the curves with embedding degrees $k \in \{3, 4, 6\}$ for almost prime order.

**GMV method**: Galbraith et al. [33] also extended the MNT method for constructing elliptic curves with embedding degree $k \in \{3, 4, 6\}$ by allowing cofactor $h \in \{2, 3, 4, 5\}$. They followed the MNT method but substituted $hr$ with $\#E(F_q)$. In the case of prime order, every parameterisation of $t(u)$ is linear, and every parameterisation of $q(u)$ is quadratic; thus, final CM equations $Dy^2 = 4q(x) - t(x)^2$ are quadratic, which enables for transforming into a generalised Pell equation.

**Freeman curve of embedding degree k=10.** Freeman [19] addressed the open problem posed by Boneh et al. [35] by constructing an elliptic curve with embedding degree 10. He chooses $n(x)$ and $t(x)$ in such a way that the high degree value of $t(x)^2$ cancel out with those of $4r(x)$ in quadratic equation $f(x) = 4r(x) - (t(x) - 2)^2$. He found that this is possible only with a curve of embedding degree 10.

To construct a curve with embedding degree k = 10, Freeman chooses following parameters:

$$r(u) = 25u^4 + 25u^3 + 15u^2 + 5u + 1$$
$$q(u) = 25u^4 + 25u^3 + 25u^2 + 10u + 3$$

We have

$$t(u) = 10u^2 + 5u + 3$$

such that

$$r(u) | \Phi_{10}(q(u))$$

and,

$$t(u)^2 + 4q(u) = -(15u^2 + 10u + 3)$$

If the norm equation $u^2 - 15Dv^2 = -20$ has a solution with $u \equiv 5 \pmod{15}$, then $\{q, r, t\}$ denotes the family of curves with embedding degree 10.

***Robert Drylo method for sparse.*** Drylo [17] also constructed the variable discriminant families using sparse families. He generalised the Brezing-Weng method to construct such families, but the method is less efficient than a method for constructing variable discriminants using a complete family.

Table 7 Sparse families with variable discriminant and embedding degree $k \in \{8,10,12\}$ by Drylo method

| |
|---|
| $k = 10, D = 1, 11, \rho = 1$ |
| $q(u) = 25u^4 + 25u^3 + 25u^2 + 10u + 3$ |
| $r(u) = 25u^4 + 25u^3 + 15u^2 + 5u + 1, \quad t(u) = 10u^2 + 5u + 3$ |
| $k = 8, D = 1, \rho = \frac{3}{2}$ |
| $q(u) = \frac{1}{576}(u^6 - 6u^5 + 7u^4 - 36u^3 + 135u^2 + 186u - 63)$ |
| $r(u) = u^4 - 2u^2 + 9, \quad t(u) = \frac{1}{12}(-u^3 + 3u^2 + 5u + 9)$ |
| $k = 12, D = 1, \rho = \frac{3}{2}$ |
| $q(u) = \frac{1}{900}(u^6 - 8u^5 + 18u^4 - 56u^3 + 202u^2 + 258u - 423)$ |
| $r(u) = u^4 - 2u^3 - 3u^2 + 4u + 13, \quad t(u) = \frac{1}{12}(-u^3 + 4u^2 + 5u + 6)$ |

***Fotiadis and Konstantinou.*** Fotiadis et al. [30] extended Drylo's method [17] for constructing the first sparse families of curves and implemented them with different embedding degrees $k \in \{5,8,10,12\}$. Fotiadis et al. [30] suggested that families $\rho(q, r, t) = 2$ provide a good balance between the size of prime order r and the size of the extension field. It also presented many numerical examples of cryptographic parameters after considering the recent method of TNFS attack on reducing the complexity of DLP in field extension of composite degree. This method constructed the sparse families of curves by adopting Lee-Park's [32] method to produce an irreducible polynomial $r(x) \in \mathbb{Q}[x]$ and trace of polynomial $t(x) \in \mathbb{Q}[x]$ in such a way that $r(x) | \Phi_k(t(x) - 1)$ on some fix embedding degree $k$. Then, it determines a non-square quadratic polynomial $g(x)$, such that the CM polynomial is $f(x) = g(x)y(x)^2$, with $p(x) \in \mathbb{Q}[x]$ using Drylo's method [17] and the rest of polynomials $q(x), y(x)$ is straightforward. Table 8 discusses the parameters of cyclotomic sparse families of curves.

Table 8 Cyclotomic sparse families of curves with distinct embedding degree and $\rho$-value

| $k$ | $\rho$ | $\varphi(k)$ | $t(x)$ | $g(x)$ | $y(x)$ |
|---|---|---|---|---|---|
| 5 | 3/2 | 4 | $x + 1$ | $3x^2 - 2x + 3$ | $-(2x^2 + 2x + 1)$ |
| 8 | 3/2 | 4 | $-x^3 + 1$ | $7x^2 - 26x + 7$ | $-\frac{1}{17}(3x^2 - x + 3)$ |
| 10 | 3/2 | 4 | $x^3 + 1$ | $3x^2 + 10x + 3$ | $\frac{1}{11}(x^2 + 3x + 1)$ |
| 10 | 3/2 | 4 | $x^3 + 1$ | $15x^2 + 50x + 15$ | $\frac{1}{55}(7x^2 - x + 7)$ |
| 7 | 5/3 | 6 | $x^5 + 1$ | $208x^2 + 375x + 208$ | $\frac{1}{71}(38x^4 - 23x^3 + 50x^2 - 23x + 38)$ |
| 9 | 5/3 | 6 | $x^5 + 1$ | $8x^2 + 35x + 8$ | $-\frac{1}{109}(x^4 - 18x^3 - 4x^2 - 18x + 1)$ |
| 14 | 5/3 | 6 | $x^5 + 1$ | $4x^2 + 5x + 4$ | $-(2x^4 - 5x^3 + 6x^2 - 5x + 2)$ |
| 18 | 5/3 | 6 | $x^5 + 1$ | $4x^2 + 9x + 4$ | $-\frac{1}{19}(3x^4 - 2x^3 - 8x^2 - 2x + 3)$ |
| 30 | 7/4 | 8 | $x^7 + 1$ | $155x^2 + 350x + 155$ | $\frac{1}{9755}(433x^6 - 293x^5 - 149x^4 + 637x^3 - 149x^2 - 293x + 433)$ |
| 10 | 2 | 4 | $x + 1$ | $15x^2 + 50x + 15$ | $-\frac{1}{19}(8x^3 - 8x^2 + 1)$ |
| 14 | 2 | 6 | $-x^2 + 1$ | $4x^2 + 5x + 4$ | $3x^5 - 4x^4 + 3x^3 - 2x + 2$ |
| 18 | 2 | 6 | $x + 1$ | $4x^2 + 9x + 4$ | $-\frac{1}{19}(7x^5 - x^4 - 6x^2 - 6x + 10)$ |
| 18 | 2 | 6 | $x + 1$ | $19x^2 + 30x + 19$ | $\frac{1}{37}(26x^5 - 14x^4 - 12x^2 - 12x^2 - 12x + 29)$ |
| 15 | 2 | 15 | $x^2 + 1$ | $3x^2 - 18x + 3$ | $\frac{1}{93}(20x^7 - 8x^6 - 22x^5 + 20x^4 + 14x^3 \mp 6x^2 + 7x - 15)$ |

| 20 | 2 | 20 | $x + 1$ | $40x^2 - 55$ | $-\dfrac{1}{505}(20x^7 + 23x^6 - 43x^5 - 4x^4 + 24x^3 + 68x^2 - 88x + 20)$ |

## 5. Attacks, Security and Key Estimation

*5.1. Recent Attacks*

The Shanks algorithm is considered the intermediate algorithm to solve the DLP for a generic Abelian group. It is enough to think about prime order groups because the Pohlig-Hellman algorithm allows us to break down the problem for composite order groups into their prime elements and reconstruct the solution using the Chinese remainder theorem. A naive brute-force attack for the discrete logarithm problem would require at most $O(p)$ operations for a group of prime order p. Shanks' Baby-Step-Giant-Step (BSGS) algorithm has two steps, each having a time/space tradeoff to reduce space-time complexity. BSGS solves the discrete logarithm problem with $O(\sqrt{p})$ operations and requires $O(\sqrt{p})$ space for the hashmap. Pollard's rho technique provides the same upper bound computation time of $O(\sqrt{p})$ without the need for a large hashmap, which can become prohibitively expensive as the order of the group increases.

*L notation*. Many best-case time complexity estimations for cryptographically hard problems, such as factoring or computing the FFDLP, are written in L-notation. It approximates the time complexity of an algorithm with an input of length $n$ when the length is allowed to grow indefinitely. Understanding how the sizes of $l$ and c affect expected computational complexity can give us a rough idea. The L-notation is represented as $L_N[l, c] \coloneqq \exp[(c + o(1))(\log N)^l (\log \log N)^{1-l}]$, where $l \in [0,1]$ is the real constant and $c > 0$. The characteristic $q = L_N[l, c]$ is medium if $\tfrac{1}{3} < l < \tfrac{2}{3}$ and large if $\tfrac{2}{3} < l \leq 1$ and a field $F_{q^k}$ is at the boundary if $l = \tfrac{2}{3}$.

The index-calculus attack, developed by Adleman in 1979 and running in $L_N\left[\tfrac{1}{2}, \sqrt{2}\right]$, was the first sub-exponential technique for integer factorisation. Pomerance (who also introduced L-notation) demonstrated that his quadratic number field sieve ran asymptotically quicker in $L_N\left[\tfrac{1}{2}, 1\right]$ time [36]pomepo. Lenstra's elliptic curve factoring technique runs in $L_N\left[\tfrac{1}{2}, \sqrt{2}\right]$, with the unusual property that the time is bounded by the lowest prime factor $q$ of $n$, rather than the number $n$ itself. Pollard, Lenstra, Lenstra, and Manasse developed the number field sieve, a technique for integer factorisation that could also solve the discrete logarithm problem in $F_{q^*}$ [37]. Their innovative strategy reduced alpha in the L-complexity with an asymptotic running time of $L_N\left[\tfrac{1}{3}, c\right]$.

The algorithm was originally made for a specific circumstance in which the input integer had to have a special form: $n = r^e \pm s$, with both $r$ and $s$ being small. This variant is known as the special number field sieve (SNFS) and has $c = \sqrt[3]{\tfrac{32}{9}}$. This works well for Mersenne numbers but not for ordinary integers. Attempts to improve this algorithm led to the formation of the generic number field sieve (GNFS), which has just a minor increase in complexity of $c = \sqrt[3]{\tfrac{32}{9}} = 1.9$. The GNFS was quickly adapted to solve discrete logarithm problems in $F_{q^*}$, but it took another ten years to adjust it for $F_{q^*}$, which was called the tower number field sieve (TNFS), which has the same complexity as the GNFS. The input has a few hundred bits for $O(\sqrt{n})$ complexity and a few thousand bits for sub-exponential complexity, making the problem look cryptographically difficult. For example, if the best-known attack takes $O(\sqrt{n})$ time, we'd need n to be 256 bits to guarantee 128 bit-strength. To provide 128-bit security, we require a modulus of 3072 bits for RSA, which can be factored in $L_N\left[\tfrac{1}{3}, 1.9\right]$ time.

A $q$-bit BN curve with a prime field of $q$ bits and an embedding degree $k$ has extension field is $q^k$ bits. For instance, 256-bit BN curves have a prime field of 256-bit and embedding degree 12, so the extension field is $q^{12}$ has 3072 bits that would have 128 bits of security. Unfortunately, as suggested by Perrin [38], BN curves need to be more secure. To Kim et al. [23], the security in the extension field has changed, and it is difficult to provide a concrete estimation. Kim et al. [23] proposed a new kind of NFS algorithm, the extended tower number field sieve (exTNFS), that reduces the complexity of solving DLP in a finite field. The exTNFS asymptotically dropped down the security level of elliptic curves for pairing. DLP for 3072-bit ($q^{12}$) is more accessible to attack than DLP for a 3072-bit prime. Recently, Barbulescu et al. [26] estimated that the BN curves, which had previously achieved 128 bits of security, dropped to around 100 bits. Menezes et al. [25] indicate that the BN curve used a 383-bits length of $q$ after applying exTNFS for achieving 128-bit of security and that of BLS12 curves as 384-bits. Kiyomura et al. [39] indicate that a bit length of $q^k$ for BLS48 curves is 27,410 bits after applying exTNFS with 572 bits of $q$.

*5.2. Security of Pairing-friendly curves*

Now that we have covered pairing and complexity estimation for asymmetric protocols. Let us combine the two problems (ECDLP and FFDLP) and estimate the security of pairing-friendly curves. Pairing-based cryptography relies on the security of the underlying pairing-friendly curves for its security. We consider FFDLP in $G_T$ and ECDLP in $G_1$ and $G_2$ since the security of most pairing-based cryptography depends on the complexity of these problems.

**ECDLP**. To solve ECDLP, the attacker must perform $O(\sqrt{r})$ curve operations using Pollard's rho algorithm to recover the required result [21], which relatively simply estimates the bit security of the EDCLP. To use ECDLP with n-bit security, we must ensure that the prime order $r$ has at least 2n-bits. This is true only if the subgroup is prime-ordered. If $r$ is composite, then the number of operations is $O(\sqrt{p})$, where $p$ is the highest prime factor of $r$, which minimises the curve's security. Since $G_1$ is often generated from points on the curve $E(F_p)$, while $F_{q^k}$ is defined on the extension field, ECDLP will take less time to solve than FFDLP because, at $E(F_p)$, the individual operations are more efficient than of $F_{q^k}$.

**FFDLP**. Compared with the group of points on an elliptic curve, we must use the structure of finite filed $F_{q^k}$ to derive the sub-exponential algorithm for solving FFDLP. The TNFS was the most famous attack for solving FFDLP. We need 3072-bit modulus p to provide 128-bit security. For pairing-friendly curves with target group $F_{q^k}$ we need 3072/k bits of field characteristic.

Additionally, if we work within the subgroup of $F_{q^k}$, the most significant prime factor must have at least 256 bits to secure against attacks such as Pollard's Rho. Thus, the security level of the pairing-friendly curve is determined by the computation complexity of an algorithm to solve the DLP on finite field $F_{q^k}$. The most effective and efficient attack on FFDLP is the index calculus method solved in sub-exponential time: $\exp[(c + o(1))(\log q^k)^{1/2}(\log \log q^k)^{1/2}]$. There have been seen several improvements over the index calculus method, such as FFS (small characteristic), and NFS (large characteristic) that solve the FFDLP in sub-exponential time: $\exp[(c + o(1))(\log q^k)^{1/3}(\log \log q^k)^{2/3}]$.

A recent study addressing the FFDLP indicates that when the extension field's characteristic $q$ or embedding degree $k$ has unique properties, the complexity decreases, as was the case for the original SNFS when a special form factoring the integers had less asymptotic complexity. The asymptotic complexity of a finite field $F_{q^k}$ is determined by its size. A pairing-based cryptosystem may be possible with $F_{q^k}$, where $q^k$ size is large. QPA, NFS-HD, and NFS are the most accepted index calculus methods for qualifying small, medium, and significant characteristics (q prime). Finite fields such as $F_{2^k}$, where $k$ is composite, are easily attacked via a quasi-polynomial algorithm, which means that curves in small characteristics, such as supersingular curves, can be broken. There may be few supersingular curves in a significant characteristic that are safe for large finite field $F_{q^k}$. Field characteristics are chosen sparsely for performance in pairing-friendly curves, and the embedding degree is often factorable, such as k=12 for both BLS12 and BN curves. This implies that the specific case complexity of SexTNFS, rather than GNFS, should be used to evaluate the bit security of various pairing-friendly curves.

*5.3. Polynomial selection*

Polynomial selection is the key to computing the average running time of an algorithm and developing a new NFS variation. Kim and Barbulescu [23] describe a new polynomial selection method that combines the TNFS strategy with the conjugation technique and improves the asymptotic complexity of the algorithm. This new version is known as exTNFS. JLSV1 [40], conjugate method [41], JLSV2 [40], modified JL [42], and Sarkar-Singh [43] are some of the polynomial selection methods for finite field $F_{q^k}$, as shown in Table 9. Usually, the NFS attacks on finite field extension with complexity $L_N[1/3,1.923]$. This complexity still works for finite extensions of prime degree. Now, there has been noticed recent progress in the TNFS method, such as extension TNFS (exTNFS) and special exTNFS (SexTNFS) approaches [23] that reduce the complexity of DLP on finite field extensions for composite $k$ and special form of $q$, to $L_N[1/3,1.526]$.

Table 9. Asymptotic complexity of different polynomial selection methods

| Polynomial selection | Characteristic size | Asymptotic complexity |
|---|---|---|
| JLSV1 [40] | Medium | $L_N[1/3,2.42]$ |
| Conjugate [41] | Medium | $L_N[1/3,2.201]$ |
| JLSV2 [40] | Large | $L_N[1/3,1.923]$ |
| Generalized JL [42] | Large | $L_N[1/3,1.923]$ |
| Sarkar-Singh [43] | Tradeoff between [42] and conjugate method [41] | $L_N[1/3,1.526]$ |
| Kim-Barbulescu [23] | Medium | $L_N[1/3,1.526]$ |

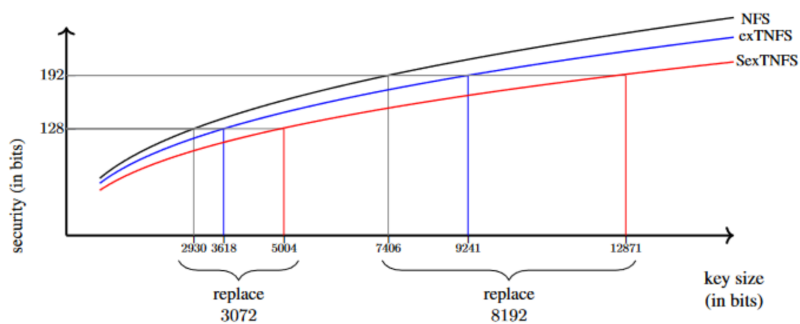

Fig. 2 Modification of key sizes according to the O(1)-less formula [26]

## 5.4. Security level and key estimation

ExTNFS with the Joux–Pierrot [44] polynomial selection for exceptional primes or exTNFS with conjugation are currently the best attacks. Table 10 shows how to estimate the key size of a few pairing-friendly elliptic curves using a rough and theoretical approach. After discovering the SexTNFS method, nearly all pairing-friendly curves previously created should be re-evaluated to estimate the DLP's new bit security [25]. They conducted experiments for the complexity 256-bit BN curve, which had previously been thought to have a security of 128 bits, among other estimations. They predicted a new complexity between 150 and 110-bit security, constrained by the value of the constants appearing in the O(1) term in $L_N[a,c]$, thanks to the improvements of the SexTNFS.

Simplification of O(1) = 0 was graphically presented by Barbulescu and Duquesne [26], allowing for the estimation of expected key sizes for pairing-friendly curves, illustrated in Fig 2. With the evolution of SexTNFS, we will need to expand the bit length of $p = q^k$ to around 5004 bits for 128bit security and 12871 bits for 192bit security. They also assert that a precise assessment of the o(1) constant term is required, particularly for the newer exTNFS and SexTNFS. They estimate that due to careful polynomial selection in the sieving process, a 256-bit BN curve will only have 100 bits of security and that other pairing-friendly curves will suffer similarly. They concluded that pairing-friendly curves with big primes should be examined for 128-bit security.

Guillevic studied the bit security of curves and published a selection of the best curves with 128-bit security for each position [28]. For efficient non-conservative pairings, Guillevic recommends BLS12-381 (or any other BLS12 curve or Fotiadis–Martindale curve of roughly 384 bits), and BLS12 or a Fotiadis–Martindale curve of 440 to 448 bits for conservative but still efficient pairings. Several novel curves guarantee 128-bit security, despite recent modifications to security estimates, and many of them can be smoothly integrated into current pairing methods. For those who already have a functional solution in a stricter space utilising BLS12-381, studies using Guillevic's sage code showed that implementing SexTNFS might reduce security to only 120 bits of security, making it a thoroughly inefficient task to deal with right now.

Table 10. Security level estimation and key sizes of few pairing-friendly elliptic curves.

| $\log q$ | $k$ | $\log p$ | Joux-Pierrot NFS [44] $d=4, L_N[1/3,2.07]$ | TNFS $L_N[1/3,1.92]$ | exTNFS $L_N[1/3,1.74]$ | SexTNFS $L_N[1/3,1.53]$ |
|---|---|---|---|---|---|---|
| 256 | 12 | 3072 | $\approx 2^{149}$ | $\approx 2^{139}$ | $\approx 2^{126}$ | $\approx 2^{110}$ |
| 384 | 12 | 4608 | $\approx 2^{177}$ | $\approx 2^{164}$ | $\approx 2^{149}$ | $\approx 2^{130}$ |
| 448 | 12 | 5376 | $\approx 2^{189}$ | $\approx 2^{175}$ | $\approx 2^{159}$ | $\approx 2^{139}$ |
| 512 | 12 | 6144 | $\approx 2^{199}$ | $\approx 2^{185}$ | $\approx 2^{168}$ | $\approx 2^{147}$ |

## 6. Pairing-friendly curves resilient to TNFS attacks

The new TNFS breakthrough substantially impacts the construction of the pairing-friendly elliptic curve with composite degree $k$. The finite field extension must be larger than the previously chosen field extension; hence $\rho = 1$ will not be the preferable choice for composite k. BN curves with k = 12, for example, were an excellent choice for generating the 256-bit prime and 3072-bit extension field with $\rho = 1$. Such parameters had previously achieved the 128-bit security level before TNFS assaults. By chance, the extension field of this size now has a 110-bit security level, thanks to the advancement of the TNFS approach. To achieve 128-bit security, the extension field should be roughly 4608 bits in size. Remember that BN curves have a value $\rho \simeq 1$, which will respond to the size of a subgroup of order $r = 384$, causing the security of $G_1, G_2$, and the target group $G_T$ to be imbalanced.

Fotiadis et al. [24] recently acknowledged the impact of TNFS variations and updated the conditions for creating polynomial families of curves $(q(x), r(x), t(x))$. They offered the best families for achieving a balanced security level in a tuple of pairing groups $(G_1, G_2, G_T)$, as well as pairing-friendly settings for composite embedding degrees that are safe from TNFS assaults. They also advised using those polynomial families of prime embedding degrees that had previously been overlooked due to the enormous ρ-value.

*Composite embedding degree.* Fotiadis et al. [24] construct complete and CVD families of curves for different composite embedding degrees that have generated the suitable curve parameters secured against the TNFS attacks. Compared to the previously discussed results, these families have a large ρ-value to expand the size of the extension field. Subsequently, it enhances the complexity of the DLP in the target group $G_T$. The variant of TNFS attacks, SexTNFS achieved the asymptotic complexity of DLP in the extension field of the composite degree to $L_N[1/3,1.526]$ where $N = q^k$.

*Prime embedding degree.* Fotiadis et al. [24] also suggested those inefficient complete and CVD families of curves for prime embedding degree that did not previously notice due to large ρ-value. Choosing curves with large ρ-value is to produce a balanced security level in three $G_1, G_2$ and $G_T$. However, the variant of TNFS attacks does not affect the prime degree extension fields, where the complexity of the DLP in is evaluated to $L_N[1/3,1.923]$, where $N = q^k$.

Table 11 Recommended family of curves at the 128-bit security level [24].

| Curves | $k$ | $D$ | $r$ bits | $q$ bits | $q^k$ bits | Seed $u$ | Security level |
|---|---|---|---|---|---|---|---|
| Cock-Pinch | 6 | 3 | 256 | 672 | 12255 | $2^{128} - 2^{124} - 2^{59}$ | 128 |
| Cock-Pinch | 8 | 1 | 256 | 544 | 13799 | $2^{64} - 2^{54} + 2^{37} + 2^{32} - 4$ | 131 |
| Cyclo FM | 10 | 15 | 256 | 446 | 12255 | $2^{32} - 2^{26} - 2^{17} + 2^{10} - 1, a = -3$ | 133 |
| Cyclo FM | 11 | 3 | 258 | 333 | 11477 | $-2^{13} + 2^{10} - 2^8 - 2^5 - 2^3 - 2, b = 13$ | 131 |
| Cyclo FM | 11 | 11 | 256 | 412 | 12255 | $-2^{56} + 2^{21} + 2^{19} - 2^{11} - 2^9 - 1, a = 2$ | 145 |
| BN | 12 | 3 | 446 | 446 | 13799 | $2^{110} + 2^{36} + 1, b = 257$ | 132 |
| Cyclo BLS | 12 | 3 | 229 | 446 | 12255 | $-2^{74} - 2^{73} - 2^{63} - 2^{57} - 2^{50} - 1, b = 1$ | 132 |
| FK | 12 | 3 | 296 | 446 | 11477 | $-2^{72} - 2^{71} - 2^{36}, b = -2$ | 136 |
| Cyclo | 13 | 3 | 267 | 310 | 12255 | $2^{11} + 2^8 - 2^6 - 2^4, b = -17$ | 140 |
| Cyclo | 14 | 3 | 256 | 340 | 13799 | $2^{21} + 2^{19} + 2^{10} - 2^6, b = -4$ | 148 |
| KSS16 | 16 | 1 | 257 | 330 | 12255 | $-2^{34} + 2^{27} - 2^{23} + 2^{20} - 2^{11} + 1, a = 1$ | 140 |
| KSS16 | 16 | 1 | 256 | 330 | 11477 | $2^{34} - 2^{30} + 2^{26} + 2^{23} + 2^{14} - 2^5 + 1, a = 1$ | 140 |

*128-bit security.* For 128-bit of security, the finite field size required for BLS12 and BN curves is about 12*448=5376. To reduce the investigation of other families of curves to those where $q^k$ is smaller than 5376 bits. To claim the curve security, the size of r should be at least 256 bits. At 128-bit security level, we decide the limit as $3072 \leq 256\rho k \leq 5376$ to reduce the number of families to consider. We can get the $k \leq 21$ and $k \geq 6$ for $\rho = 1$ and $\rho = 2$, respectively. Table 11 summarises the final SNTFS secure pairing-friendly elliptic curve with embedding degree $k = \{6,8,10,11,12,13,14,16\}$.

*192-bit security.* Guillevic et al. [28] assumed the limit $7168 \leq 384\rho k \leq 14336$ for 192-bit level of security. We can get $k \leq 37$ and $k \geq 10$ for $\rho = 1$ and $\rho = 2$, respectively. Fotiadis et al. [24] curves with $\rho = 2$ satisfy the boundary for $10 \leq k \leq 18$. There could not found any cyclotomic family of curves with embedding degree $k = 32$. Table 12 shows the recommended parameters for families of curves at a 192-bit level of security.

Table 12. Recommended family of curves at the 192-bit security level [24].

| Curves | $k$ | $r$ bits | $p$ bits | $p^k$ bits | Seed $u$ | Security level |
|---|---|---|---|---|---|---|
| BN | 12 | 1024 | 1022 | 12255 | $-2^{254} + 2^{33} + 2^6$ | 191 |
| BLS12 | 12 | 768 | 1150 | 13799 | $-2^{192} + 2^{188} - 2^{115} - 2^{110} - 2^{44} - 1$ | 193 |
| KSS16 | 16 | 605 | 766 | 12255 | $2^{78} - 2^{76} - 2^{28} + 2^{14} + 2^7 + 1$ | 194 |
| KSS18 | 18 | 474 | 638 | 11477 | $2^{80} + 2^{77} + 2^{76} - 2^{61} - 2^{53} - 2^{14}$ | 193 |
| BLS24 | 24 | 409 | 509 | 12202 | $-2^{51} - 2^{28} + 2^{11} - 1$ | 193 |

## 7. Discussion and Recommendation

We have a special NFS algorithm [23] for prime fields, where $q$ has a special form, whose complexity is $L_N\left[\frac{1}{3}, (\frac{32}{9})^{\frac{1}{3}} = 1.526\right]$. For example, Joux and Pierrot [44] demonstrate a dedicated polynomial selection method for extension fields $F_{q^k}$, where a polynomial of degree $d$ derives q at given value (d = 4 for BN curves), such that NFS algorithm has a different expected running time for different characteristics. In this way, the asymptotic complexity of NFS algorithm to attack the target group of pairing has been reduced. We recommend that DLP over medium characteristic fields are so harder than large characteristic fields. Previously, the recommended size of $F_{q^k}$ was evaluated by assuming asymptotic complexity $L_N\left[\frac{1}{3}, 1.923\right]$. As the new complexity is below this bound, the size should be enlarged. Kim and Barbulescu [23] obtained the smallest asymptotic complexity, that is $L_N\left[\frac{1}{3}, (\frac{32}{9})^{\frac{1}{3}} = 1.526\right]$ in the case of medium characteristics by combining the Joux-Pierrot method with TNFS method. Thus, the complexity of exTNFS does not depend on the characteristic size

Improvements in the SexTNFS's asymptotic complexity do not weaken current pairing-based protocols considerably enough to trigger significant concern over protocol changes. The protocol is now secure against all feasible attacks, thanks to the reduction in security from 128 bits to 120 bits. Nonetheless, it changes how we discuss the bit security of specific pairing-friendly curves. Another security flaw with less computing power may be uncovered before a pairing-based protocol with "just" 120 or even 100-bit security fails. The emergence of quantum computers, which, like other cryptographic protocols that rely on the rigour of the discrete logarithm problem for some abelian group, break pairing-based protocols, is a more prominent issue. To ensure that our protocols perform securely, we must understand the innovative advances affecting pairing-based protocols. Famous curves, such as BLS12-381, are still secure and pairing-friendly, but they are not 128-bit secure pairing curves. We will need the newly produced curves if we need security at a specific NIST level or declare that a protocol implementation is 128-bit secure. We should stay updated and use new, fine-tuned curves like the Fotiadis-Martindale Curve [29] or the better BLS Curve BLS12-440 [26].

We observe that out of all computationally difficult problems, the discrete log problem using TNFS has the most significant disparity between the computational record and the security estimates. Due to the pairing-based protocol, computing the FFDLP has become popular recently. The most powerful attack, even in pairing-friendly curves, employs the known Pollard's rho technique for solving the discrete logarithm problem on elliptic curves rather than on (sex) TNFS at the output of the pairing. Modern breakthroughs have led cryptographers to create new 128-bit secure curves. However, the earlier curves found in BLS 12-381 are still excellent choices for cryptographic protocols. Their use of smaller primes has led to their 440-bit bit being significantly faster than their counterparts.

## 8. Conclusion

The topic of choosing secure pairing-friendly curves has reappeared because of recent attacks on discrete logs in $F_{q^k}$. Many families of pairing-friendly elliptic curves have recently been discovered vulnerable to number field sieve (NFS) attacks. The problem of selecting pairing-friendly curves at various security levels is explored in this study. First, we added to Freeman's classification by recognising new families that had not been mentioned before, such as an entire family with variable differentiation and new sparse curve families. We have investigated a comprehensive framework for constructing an individual as well as parameterise families of curves. We have assessed the operational security of various families of pairing-friendly curves in terms of required key size to uncover better families of curves than BN, KSS, and BLS. We have illustrated that recent attacks, such as TNFS on pairing, need to enhance the key size for pairing, so we analysed families of curves in terms of their key size and selected an appropriate choice of an elliptic curve. Further, we recommend some better and more secure pairing-friendly curves.

## References

[1] D. Freeman, M. Scott, and E. Teske, "A taxonomy of pairing-friendly elliptic curves," *J. Cryptol.*, vol. 23, no. 2, pp. 224–280, 2010.
[2] R. Barbulescu, N. El Mrabet, and L. Ghammam, "A taxonomy of pairings, their security, their complexity," 2019.
[3] W. Diffie and M. Hellman, "New directions in cryptography," *IEEE Trans. Inf. Theory*, vol. 22, no. 6, pp. 644–654, 1976.
[4] N. Koblitz, "Elliptic curve cryptosystems," *Math. Comput.*, vol. 48, no. 177, pp. 203–209, 1987.
[5] S. MilierV, "Use ofelliptic curve in cryptography," *Advannce in Cryptology—CRYPTO*, vol. 85, pp. 417–426.
[6] A. J. Menezes, T. Okamoto, and S. A. Vanstone, "Reducing elliptic curve logarithms to logarithms in a finite field," *iEEE Trans. Inf. Theory*, vol. 39, no. 5, pp. 1639–1646, 1993.
[7] G. Frey and H.-G. Rück, "A remark concerning *m*-divisibility and the discrete logarithm in the divisor class group of curves," *Math. Comput.*, vol. 62, no. 206, pp. 865–874, 1994.
[8] P. S. L. M. Barreto and M. Naehrig, "Pairing-friendly elliptic curves of prime order," in *International Workshop on Selected Areas in Cryptography*, 2005, pp. 319–331.
[9] A. J. Menezes, *Elliptic Curve Public Key Cryptosystems*, vol. 234. Springer Science & Business Media, 1993.
[10] A. Miyaji, M. Nakabayashi, and S. Takano, "New explicit conditions of elliptic curve traces for FR-reduction," *IEICE Trans. Fundam. Electron. Commun. Comput. Sci.*, vol. 84, no. 5, pp. 1234–1243, 2001.
[11] P. S. L. M. Barreto, B. Lynn, and M. Scott, "On the selection of pairing-friendly groups," in *International Workshop on Selected Areas in Cryptography*, 2003, pp. 17–25.
[12] R. Dupont, A. Enge, and F. Morain, "Building curves with arbitrary small MOV degree over finite prime fields," *J. Cryptol.*, vol. 18, no. 2, pp. 79–89, 2005.
[13] C. Cocks and R. G. E. Pinch, "Identity-based cryptosystems based on the Weil pairing (2001)," *Unpubl. Manuscr.*
[14] F. Brezing and A. Weng, "Elliptic curves suitable for pairing based cryptography," *Des. Codes Cryptogr.*, vol. 37, no. 1, pp. 133–141, 2005.
[15] E. J. Kachisa, E. F. Schaefer, and M. Scott, "Constructing Brezing-Weng pairing-friendly elliptic curves using elements in the cyclotomic field," in *International Conference on Pairing-Based Cryptography*, 2008, pp. 126–135.
[16] S. Tanaka and K. Nakamula, "Constructing pairing-friendly elliptic curves using factorisation of cyclotomic polynomials," in *International Conference on Pairing-Based Cryptography*, 2008, pp. 136–145.
[17] R. Dryło, "On constructing families of pairing-friendly elliptic curves with variable discriminant," in *International conference on cryptology in india*, 2011, pp. 310–319.
[18] M. Scott and P. S. L. M. Barreto, "Generating more MNT elliptic curves," *Des. Codes Cryptogr.*, vol. 38, no. 2, pp. 209–217, 2006.
[19] D. Freeman, "Constructing pairing-friendly elliptic curves with embedding degree 10," in *International Algorithmic Number Theory Symposium*, 2006, pp. 452–465.

[20] M. Scott and A. Guillevic, "A new family of pairing-friendly elliptic curves," in *International Workshop on the Arithmetic of Finite Fields*, 2018, pp. 43–57.

[21] J. M. Pollard, "Monte Carlo methods for index computation (modp)," *Math. Comput.*, vol. 32, no. 143, pp. 918–924, 1978.

[22] J. Jeong and T. Kim, "Extended Tower Number Field Sieve with Application to Finite Fields of Arbitrary Composite Extension Degree.," *IACR Cryptol. ePrint Arch.*, vol. 2016, p. 526, 2016.

[23] T. Kim and R. Barbulescu, "Extended tower number field sieve: A new complexity for the medium prime case," in *Annual International Cryptology Conference*, 2016, pp. 543–571.

[24] G. Fotiadis and E. Konstantinou, "TNFS resistant families of pairing-friendly elliptic curves," *Theor. Comput. Sci.*, vol. 800, pp. 73–89, 2019.

[25] A. Menezes, P. Sarkar, and S. Singh, "Challenges with assessing the impact of NFS advances on the security of pairing-based cryptography," in *International Conference on Cryptology in Malaysia*, 2016, pp. 83–108.

[26] R. Barbulescu and S. Duquesne, "Updating key size estimations for pairings," *J. Cryptol.*, vol. 32, no. 4, pp. 1298–1336, 2019.

[27] A. Guillevic and S. Singh, "On the alpha value of polynomials in the tower number field sieve algorithm," 2019.

[28] A. Guillevic, "A short-list of pairing-friendly curves resistant to Special TNFS at the 128-bit security level," in *IACR International Conference on Public-Key Cryptography*, 2020, pp. 535–564.

[29] G. Fotiadis and C. Martindale, "Optimal TNFS-secure pairings on elliptic curves with even embedding degree," 2018.

[30] G. Fotiadis and E. Konstantinou, "Generating pairing-friendly elliptic curve parameters using sparse families," *J. Math. Cryptol.*, vol. 12, no. 2, pp. 83–99, 2018.

[31] P. S. L. M. Barreto, B. Lynn, and M. Scott, "Constructing elliptic curves with prescribed embedding degrees," in *International Conference on Security in Communication Networks*, 2002, pp. 257–267.

[32] H.-S. Lee and C.-M. Park, "Generating pairing-friendly curves with the CM equation of degree 1," in *International Conference on Pairing-Based Cryptography*, 2009, pp. 66–77.

[33] S. D. Galbraith, J. F. McKee, and P. C. Valença, "Ordinary abelian varieties having small embedding degree," *Finite Fields Their Appl.*, vol. 13, no. 4, pp. 800–814, 2007.

[34] Y. Nanjo, M. Shirase, Y. Kodera, T. Kusaka, and Y. Nogami, "A Construction Method of Final Exponentiation for a Specific Cyclotomic Family of Pairing-Friendly Elliptic Curves with Prime Embedding Degrees," in *2021 Ninth International Symposium on Computing and Networking (CANDAR)*, 2021, pp. 148–154.

[35] D. Boneh, B. Lynn, and H. Shacham, "Short signatures from the Weil pairing," in *International Conference on the Theory and Application of Cryptology and Information Security*, 2001, pp. 514–532.

[36] C. Pomerance, "On composite n for which φ (n)| n− 1. II," *Pacific J. Math.*, vol. 69, no. 1, pp. 177–186, 1977.

[37] A. K. Lenstra, H. W. Lenstra Jr, M. S. Manasse, and J. M. Pollard, "The number field sieve," in *Proceedings of the twenty-second annual ACM symposium on Theory of computing*, 1990, pp. 564–572.

[38] T. Perrin, "Curves for pairings," 2016. https://moderncrypto.org/mail-archive/curves/2016/000740.html (accessed Mar. 07, 2022).

[39] Y. Kiyomura, A. Inoue, Y. Kawahara, M. Yasuda, T. Takagi, and T. Kobayashi, "Secure and efficient pairing at 256-bit security level," in *International Conference on Applied Cryptography and Network Security*, 2017, pp. 59–79.

[40] A. Joux, R. Lercier, N. Smart, and F. Vercauteren, "The number field sieve in the medium prime case," in *Annual International Cryptology Conference*, 2006, pp. 326–344.

[41] R. Barbulescu, P. Gaudry, A. Guillevic, and F. Morain, "Improving NFS for the discrete logarithm problem in non-prime finite fields," in *Annual International Conference on the Theory and Applications of Cryptographic Techniques*, 2015, pp. 129–155.

[42] A. Joux and R. Lercier, "Improvements to the general number field sieve for discrete logarithms in prime fields. A comparison with the Gaussian integer method," *Math. Comput.*, vol. 72, no. 242, pp. 953–967, 2003.

[43] P. Sarkar and S. Singh, "New complexity tradeoffs for the (multiple) number field sieve algorithm in non-prime fields," in *Annual International Conference on the Theory and Applications of Cryptographic Techniques*, 2016, pp. 429–458.

[44] A. Joux and C. Pierrot, "The Special Number Field Sieve in $\mathbb{F}_{p^{n}}$," in *International Conference on Pairing-Based Cryptography*, 2013, pp. 45–61.